\documentstyle[12 pt]{article}

\begin{document}
\begin{center}
\bigskip
{\bf {\Large Quantum mechanics without spacetime III}}\\
\smallskip
{\bf - A proposal for a non-linear Schrodinger equation -}

\bigskip
\bigskip

{\bf {\large 
T. P. Singh\footnote{e-mail address: tpsingh@tifr.res.in}}}

{\it Tata Institute of Fundamental Research,}\\
{\it Homi Bhabha Road, Mumbai 400 005, India}

\end{center}

\bigskip
\bigskip

\begin{abstract}

\noindent It is proposed that the Schrodinger equation for a free point particle has non-linear corrections which depend on the mass of the particle. It is assumed that the corrections become extremely small when the mass is much smaller or much larger than a critical value (the critical value being related to but
smaller than Planck mass). The corrections become significant when the mass is close to this critical value and could play a role in explaining wave-function collapse. It appears that such corrections are not ruled out by present day experimental tests of the Schrodinger equation. Corrections to the energy levels of a harmonic oscillator are calculated.
\end{abstract}

\def\singlespace {\smallskipamount=3.75pt plus1pt minus1pt
                  \medskipamount=7.5pt plus2pt minus2pt
                  \bigskipamount=15pt plus4pt minus4pt
                  \normalbaselineskip=15pt plus0pt minus0pt
                  \normallineskip=1pt
                  \normallineskiplimit=0pt
                  \jot=3.75pt
                  {\def\smallskip {\vskip\smallskipamount}}
                  {\def\medskip   {\vskip\medskipamount}}
                  {\def\bigskip   {\vskip\bigskipamount}}
                  {\setbox\strutbox=\hbox{\vrule
                    height10.5pt depth4.5pt width 0pt}}
                  \parskip 7.5pt
                  \normalbaselines}

\section{Introduction}
The fact that Planck mass, $m_{p}=(\hbar c/G)^{1/2}\sim 10^{-5}$ gms, is much larger than elementary particle masses, has sometimes led to the suggestion \cite{fey} that standard, linear quantum mechanics may not hold for idealized point particles whose mass approaches the Planck mass. In particular, it is possible that the Schrodinger equation for the free particle may have to be modified to include non-linear terms, when the particle's mass becomes comparable to $m_{p}$. This is the possibility considered in this brief note.

From the experimental viewpoint, Planck mass is however, embarrasingly, in the macroscopic mass range, and its more likely that the fundamental mass scale $m_{c}$ where any non-linearities might set in is a few orders of magnitude smaller than $m_{p}$. The reason how this happens is not clear, though it is plausible that an underlying theory relates $m_{c}$ directly to $m_{p}$.

Even if $m_{c}$ is a few orders of magnitude smaller than Planck mass, there is still a very huge gap between elementary particle masses and $m_{c}$. It is significant that quantum mechanics has not been experimentally tested for a point particle with a mass $m$ somewhere in this range, say $m=10^{-15}$ gms. The purpose of this note is to give one example of a mass-dependent non-linear modification of the Schrodinger equation, with a view to encouraging experimental searches of a possible mass-dependent non-linearity.
  
Considerations of non-linear generalizations of the Schrodinger equation have a long history, an early work being that of Bialynicki-Birula and Mycielski \cite{bia}; more recent discussions include those by Weinberg \cite{wei}, Zeilinger \cite{zei}, Pearle \cite{pea}, Shimony \cite{shi} and Minic et al. \cite{min}. A more complete list of references can be found in the recent work of Adler \cite{adl}. A mass dependent non-linearity does not seem to have been considered before, though.

\section{A non-linear Schrodinger equation}

Let us rewrite the free-particle Schrodinger equation 
\begin{equation}
i\hbar{\partial\psi\over\partial t}= 
-{\hbar^{2}\over 2m} {\partial^{2}\psi\over\partial q^{2}}
\label{sch}
\end{equation}
by defining $\psi=e^{iS(q,t)/\hbar}$, where $S(q,t)$ is complex. This gives the following equation for $S(q,t)$:
\begin{equation} 
{\partial S\over\partial t} = -{1\over 2m}
\left(\partial S\over\partial q\right)^{2} +
{i\hbar\over 2m} {\partial^{2}S\over\partial q^{2}}.
\label{ess}
\end{equation}
In the so-called $\hbar\rightarrow 0$ limit the last term of this equation is dropped, and one recovers the classical Hamilton-Jacobi equation for the action function $S(q,t)$.

Let us now assume that Eqn. (\ref{ess}) has mass-dependent corrections - these corrections are assumed to be vanishingly small for masses which are much smaller or much larger than the fundamental mass scale $m_{c}$, but become significant when $m$ is comparable to $m_{c}$.

It is obvious that the equation (\ref{ess}) for the generalized action $S$ is non-linear, the non-linearity being introduced by the term quadratic in $\partial S/\partial q$. We assume that the proposed mass-dependent modification of this equation is such that precisely for $m=m_{c}$ equation (\ref{ess}) actually becomes linear! As $m$ moves away from $m_{c}$ (becoming either larger or smaller than $m_{c}$) the non-linear term in (\ref{ess}) is recovered. Thus, we propose the following modified equation, by introducing the modifying function $A(m/m_{c})$:
\begin{equation} 
{\partial S\over\partial t} = {1\over 2m}\left(-1+A\left({m\over m_{c}}\right)\right)
\left(\partial S\over\partial q\right)^{2} +
{i\hbar\over 2m} {\partial^{2}S\over\partial q^{2}}.
\label{mod}
\end{equation}

The function $A(m/m_{c})$ takes the value one at $m=m_{c}$ and goes to zero when $m$ is much larger or much smaller than $m_{c}$. Hence, for $m=m_{c}$ the modified equation for $S(q,t)$ is 
\begin{equation}
i\hbar{\partial S\over\partial t}= 
-{\hbar^{2}\over 2m} {\partial^{2} S\over\partial q^{2}}
\label{lins}
\end{equation}
which has the same form as the Schrodinger equation!

In Eqn. (\ref{mod}), after substituting for $S$ in terms of $\psi$, one gets the following non-linear Schrodinger equation for $\psi(q,t)$:
\begin{equation}
i\hbar{\partial\psi\over\partial t}= 
-{\hbar^{2}\over 2m} {\partial^{2}\psi\over\partial q^{2}}
+{\hbar^{2}\over 2m}A(m/m_{c})\left({\partial\ln\psi\over\partial q}\right)^{2}\psi.
\label{modse}
\end{equation}

Consider the implication of this non-linearity for the special case $m=m_{c}$.
Let $\psi_{1}$ and $\psi_{2}$ be two solutions of the Schrodinger equation 
(\ref{sch}), so that the superposition $\psi=a\psi_{1}+b\psi_{2}$ also solves (\ref{sch}). Since (\ref{lins}) has the same structure as (\ref{sch}), one can choose solutions for the generalised action $S$ in (\ref{lins}) as $S_{1}=\psi_{1}$ and $S_{2}=\psi_{2}$. The non-linear Schrodinger equation thus has two solutions $\psi_{1NL}=e^{i\psi_{1}/\hbar}$ and $\psi_{2NL}=e^{i\psi_{2}/\hbar}$ which have the same information content as $\psi_{1}$ and $\psi_{2}$ but the superposition of $\psi_{1NL}$ and $\psi_{2NL}$ does not satisfy the non-linear equation (\ref{modse}). If one regards the quantum measurement process as a sudden jump in the mass of a system from a value $m\ll m_{c}$ to a value $m\gg m_{c}$, the breakdown in superposition induced by the non-linearity could possibly explain wave-function collapse. This is because during the measurement $m$ `passes` through the intermediate value $m_{c}$. Here, before the measurement, by system one means the quantum system (hence $m\ll m_{c}$) and after the measurement by system one means [measuring apparatus + quantum system] (hence $m\gg m_{c}$). 

A plane wave $\psi=\alpha\exp -i(\omega t-kq)$ satisfies the non-linear equation (\ref{modse}) with the dispersion relation
\begin{equation}
\hbar\omega={\hbar^{2}\over 2m}k^{2}(1-A)
\end{equation}
but a superposition of plane-waves does not satisfy this equation. 

In Eqn. (\ref{mod}) let us make the substitutions
\begin{equation}
\tilde{q}=\frac{q}{\sqrt{1-A(m/m_{c})}}, \quad \hbar_{N}=
\frac{\hbar}{1-A(m/m_{c})}.
\label{redef}
\end{equation}
This changes (\ref{mod}) to the form
\begin{equation}
{\partial S\over\partial t} = -{1\over 2m}
\left({\partial S\over\partial \tilde{q}}\right)^{2} +
{i\hbar_{N}\over 2m} {\partial^{2}S\over\partial \tilde{q}^{2}}.
\label{mod2}
\end{equation}

This has the same form as (\ref{ess}), with $\tilde{q}$ replacing $q$, and $\hbar_{N}$ replacing $\hbar$. It is evident that if we now define $\psi_{N}=
e^{iS/\hbar_{N}}$ then (\ref{mod2}) is transformed to the following linear equation
\begin{equation}
i\hbar_{N}{\partial\psi_{N}\over\partial t}= 
-{\hbar_{N}^{2}\over 2m} {\partial^{2}\psi_{N}\over\partial \tilde{q}^{2}}
\label{sch2}
\end{equation}
which has the same form as the Schrodinger equation, with $\hbar_{N}$ replacing $\hbar$, and $\tilde{q}$ replacing $q$. It is also easily seen that $\psi_{N}$
is related to the original wave-function $\psi$ as
\begin{equation}
\psi_{N}=e^{iS/\hbar_{N}}=\left(e^{iS/\hbar}\right)^{\hbar/\hbar_{N}}=
\psi^{1-A(m/m_{c})}.
\label{red}
\end{equation}

Equivalently, one can verify that the non-linear equation (\ref{modse}) reduces to the linear equation (\ref{sch2}) after the substitutions given in (\ref{redef}) and (\ref{red}) are made. The wave-function $\psi_{N}$ satisfies the continuity equation, with the modified Planck's constant $\hbar_{N}$ replacing $\hbar$, and with $\tilde{q}$ replacing $q$. In the limit that $A$ goes to zero, $\psi_{N}$, $\hbar_{N}$ and $\tilde{q}$ reduce to $\psi$, $\hbar$ and $q$ respectively.

There appears to be a certain kind of singularity at $m=m_{c}$ which is possibly of fundamental importance.

\section{Corrections to energy levels of the harmonic oscillator}
The Schrodinger equation for the harmonic oscillator
\begin{equation}
i\hbar{\partial\psi\over\partial t}= 
-{\hbar^{2}\over 2m} {\partial^{2}\psi\over\partial q^{2}}
+{1\over 2}m\omega^{2}q^{2}\psi
\label{scho}
\end{equation}
is transformed, after the substitution $\psi=e^{iS/\hbar}$, to
\begin{equation} 
{\partial S\over\partial t} = -{1\over 2m}
\left(\partial S\over\partial q\right)^{2} +
{i\hbar\over 2m} {\partial^{2}S\over\partial q^{2}}-{1\over 2}m\omega^{2}q^{2}. \label{esso}
\end{equation}

Including the proposed mass-dependent correction, as in (\ref{mod}), changes this equation to 
\begin{equation} 
{\partial S\over\partial t} = {1\over 2m}\left(-1+A\left({m\over m_{c}}\right)\right)
\left(\partial S\over\partial q\right)^{2} +
{i\hbar\over 2m} {\partial^{2}S\over\partial q^{2}}-{1\over 2}m\omega^{2}q^{2}. \label{modo}
\end{equation}
The transformations (\ref{redef}) along with the re-definition 
$\tilde{\omega}=\omega\sqrt{1-A}$ give
\begin{equation}
{\partial S\over\partial t} = -{1\over 2m}
\left({\partial S\over\partial \tilde{q}}\right)^{2} +
{i\hbar_{N}\over 2m} {\partial^{2}S\over\partial \tilde{q}^{2}}
-{1\over 2}m\tilde{\omega}^{2}\tilde{q}^{2}.
\label{modos}
\end{equation}

The substitution $\psi_{N}=e^{iS/\hbar_{N}}$ then gives 
\begin{equation}
i\hbar_{N}{\partial\psi_{N}\over\partial t}= 
-{\hbar_{N}^{2}\over 2m} {\partial^{2}\psi_{N}\over\partial \tilde{q}^{2}}
+{1\over 2}m\tilde{\omega}^{2}\tilde{q}^{2}\psi_{N}.
\label{scho2}
\end{equation}
Noting that the time-dependent part of a stationary state 
$\psi$ goes as $e^{-iEt/\hbar}$ we can see that
\begin{equation}
i\hbar_{N}{\partial\psi_{N}\over\partial t}=E\psi_{N}
\end{equation}
and hence that the non-linear correction modifies the energy levels to
\begin{equation}
E=(n+1/2)\hbar_{N}\tilde{\omega}=(n+1/2){\hbar\omega\over\sqrt{1-A}}.
\label{osc}
\end{equation}

\section{Discussion}
The particular proposal for non-linearity given here is of course tentative; the main purpose of this note being to highlight the possibility of a mass-dependent non-linear correction, which should be probed by experiments.

The possibility of such corrections is also suggested by earlier work \cite{sin} on a spacetime-independent formulation of quantum mechanics. The structure of the formulation for a free particle suggests that if the mass of the particle is close to $m_{c}$ there should be non-linear corrections to the Schrodinger equation. The exact nature of these non-linear terms (for instance the functional form of $A(m/m_{c})$) can only be known when a better understanding of spacetime-independent quantum mechanics is achieved. Furthermore, as discussed in \cite{sin}, one expects the introduction of non-linearities to be accompanied by a change in the fundamental commutation relations.

It is known \cite{Gis} that non-linearities in a deterministic modification of the Schrodinger equation can allow for superluminal communication, and this has sometimes been presented as an argument against such a modification. But one could well speculate that such a radical change in our basic ideas might actually be occurring in the experimentally unprobed domain around $m=m_{c}$. Physicists are not averse to the possibility of a modification in quantum mechanics at the Planck energy scale $E_{p}\sim 10^{19}$ GeV. By virtue of the equivalence between mass and energy, the possibility of such a modification should also be entertained around the Planck mass scale $m_{p}=E_{p}/c^{2}$.

I would like to thank C. S. Unnikrishnan for useful discussions which lead to a significant modification of an earlier version of the paper.


\begin{thebibliography}{99}
\bibitem{fey}R. P. Feynman, Lectures on Gravitation (1962), Section 1.4
\bibitem{bia}I. Bialynicki-Birula and J. Mycielski, Ann. Phys. 100, 62 (1976)
\bibitem{wei}S. Weinberg, Phys. Rev. Lett. 62, 485 (1989)
\bibitem{zei}A. Zeilinger in {\it Quantum Concepts in Space and Time} Eds. R. Penrose and C. J. Isham, Oxford (1986)
\bibitem{pea} P. Pearle in {\it Quantum Concepts in Space and Time}  Eds. R. Penrose and C. J. Isham, Oxford (1986)
\bibitem{shi}A. Shimony in {\it Quantum Concepts in Space and Time}  Eds. R. Penrose and C. J. Isham, Oxford (1986)
\bibitem{min}D. Minic and C. H. Tze, hep-th/0305193
\bibitem{adl}S. L. Adler, Phys. Rev. D 67, 025007 (2003)
\bibitem{sin}T. P. Singh, gr-qc/0112119, to appear in {\it Proc. of the Workshop on Mach's Principle}, IIT Kharagpur, 2002, Eds. A. R. Roy and Mendel Sachs; Gen. Rel. Grav. 35, 869 (2003) [gr-qc/0205056]. 
\bibitem{Gis}N. Gisin, Helv. Phys. Acta 62, 363 (1989), Phys. Lett. A 143, 1 (1990), J. Phys. A 28, 7375 (1995); J. Polchinski, Phys. Rev. Lett. 66, 397 (1991). 
\end{thebibliography}
\end{document}